# Ultra-sensitive voltage-controlled skyrmion-based spintronic diode

Davi Rodrigues[1], Riccardo Tomasello[1], Giulio Siracusano[2], Mario Carpentieri[1], Giovanni Finocchio[3]

[1]Department of Electrical and Information Engineering, Politecnico of Bari, I-70125 Bari, Italy

[2]Department of Electric, Electronic and Computer Engineering, University of Catania, I-95125 Catania, Italy

[3]Department of Mathematical and Computer Sciences, Physical Sciences and Earth Sciences, University of Messina, I-98166, Messina, Italy

[*]corresponding authors: davi.rodrigues@poliba.it, gfinocchio@unime.it,

**Abstract**

We have designed a passive spintronic diode based on a single skyrmion stabilized in a magnetic tunnel junction and studied its dynamics induced by voltage-controlled anisotropy (VCMA) and Dzyaloshinskii-Moriya interaction (VDMI). We have demonstrated that the sensitivity (rectified voltage over input microwave power) with realistic physical parameters and geometry can be larger than 10 kV/W which is one order of magnitude better than diodes employing a uniform ferromagnetic state. Our numerical and analytical results on the VCMA and VDMI-driven resonant excitation of skyrmions beyond the linear regime reveal a frequency dependence on the amplitude and no efficient parametric resonance. Skyrmions with a smaller radius produced higher sensitivities, demonstrating the efficient scalability of skyrmion-based spintronic diodes. These results pave the way for designing passive ultra-sensitive and energy efficient skyrmion-based microwave detectors.

Spintronic diodes (SDs) leverage resonant magnetic excitations for developing high performance detectors for applications in Internet-of-Things, energy harvesting, and artificial intelligence.[1,2] Those devices implemented with magnetic tunnel junctions (MTJs) are based on the spintronic diode effect which generates a rectified voltage as a response to microwave currents.[1,2] Two key metrics of SDs are the sensitivity, and the frequency tunability.

Current sensitivity of passive SDs has reached 1 kV/W[3] for a uniform ferromagnetic state, but it can be improved to 200 kV/W[4] in SDs working in active regime, and to 4 MV/W[5] when coupled with bolometric effect. The use of alternating voltage-controlled magnetic anisotropy (VCMA) for exciting the magnetization dynamics has shown to be a key ingredient, together with the injection locking, to increase the sensitivity of SDs while also lowering Joule losses.[4,6,7] Another mean to increase the sensitivity and tunability of SDs is to leverage non-collinear magnetic textures such as magnetic vortices and skyrmions.[8–11] Studies of active vortex-based SDs have already shown sensitivities in the GHz range of up to 80 kV/W.[9] However, magnetic vortices are non-local textures and their gyrotropic dynamics are highly affected by the pinning distribution. This can be avoided by employing skyrmions, since they are localized non-collinear textures,[12–16] and the spin-diode effect can be linked directly to its breathing mode in MTJs with perpendicular polarizer.[8]

Magnetic skyrmions have been electrically detected[17,18] and their excitation modes significantly enrich their range of applications.[14,19–23] Moreover, the stability and dynamics of magnetic skyrmions are highly influenced by the Dzyaloshinskii-Moriya interaction (DMI) which can be controlled by voltage,[24–26] strain,[27] chemisorption[28] and temperature.[5] Voltage-controlled DMI (VDMI) opens new paths for skyrmion-based SDs with ultralow Joule losses. Currently predicted values for VCMA and VDMI show an efficiency of about 1 pJ/Vm.[24–26,29]

Here, we perform a systematic study of the excitation of the skyrmion breathing mode driven by VCMA and VDMI in a passive SD. The main results are that this excitation driven by VCMA is more efficient than by VDMI, with the efficiency depending on the skyrmion size. Smaller skyrmions reveal a higher sensitivity to the variations of the material parameters. We also notice that VDMI and VCMA variations induce excitations with a relative phase of $\pi/2$. This hinders the amplitude growth at resonance if both VCMA and VDMI are applied simultaneously and in phase.

We have also identified a nonlinear rectification response due to the asymmetric breathing mode (see Supplemental Material), which reveals a dependence of the frequency with the amplitude of the excitation. Numerical and analytical calculations show that excitation by fractions of the resonance frequency is more efficient than by twice of the applied frequency.[14] Hence, we show that, within a realistic set of parameters, parametric resonance of a single skyrmion is not observed. Finally, we have compared the efficiency of skyrmion-based SDs with single-domain-based SDs. We show that employing skyrmions in the MTJ can increase the sensitivity of the SD by at least one order of magnitude (given a set of parameters).

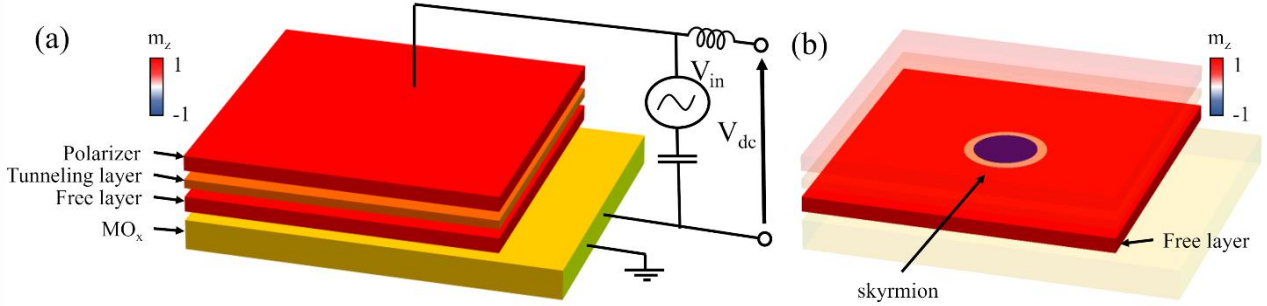

**FIG. 1.** A sketch of the stack proposed for the realization of the voltage controlled skyrmion-based detector in an MTJ. (a) Structure of the device with the indication of the different layers of the MTJ, a polarizer, a tunnelling layer, a free layer, and a metal oxide (MO$_x$) layer. A schematics of the electrical circuit to excite the skyrmion dynamics and detect the output dc voltage is also included. (b) Configuration of a skyrmion stabilized in the free layer magnetization.

*Device and numerical model.* We consider a perpendicular MTJ (see Fig. 1) composed of an iron-rich CoFeB free layer and an MgO tunneling layer, similarly to Ref.[8]. The MgO tunneling layer allows to generate the VCMA effect and control the anisotropy of the free layer. On the bottom of the free layer, we also consider a layer of a metal oxide (MO$_x$) necessary to attain a VDMI effect, induced in the free layer magnetization as discussed in the literature.[24–26,29–32] Therefore, we can employ VCMA or VDMI independently. The VCMA or VDMI effect can be achieved by controlling the relative oxygen levels on the tunneling layer or metal oxide layer, respectively, and the free layer with an applied voltage. Different specific materials and configurations of the tunneling layer and metal oxide layer have been considered in the literature to optimize the VCMA and VDMI.[7,24–27,29,30,32,33]. For the device parameters, we consider the values from Ref.[8] (see Table 1). The minimum and maximum resistance configurations of the MTJ correspond to the magnetization of the free layer homogenously parallel ($R_P$) or anti-parallel ($R_{AP}$) to the perpendicular magnetization of the polarizer. The magnetoresistance of the skyrmion state corresponds to an intermediate value between $R_P$ and anti-parallel $R_{AP}$ according to the skyrmion radius, where the core of the skyrmion is anti-parallel to the polarizer magnetization.[17] Both VCMA and VDMI can induce a change in the skyrmion radius and, hence, in the magnetoresistive value, which is converted into an dc output voltage V$_{dc}$. All numerical calculations of the magnetization dynamics are based on full micromagnetic simulations and are described by the Landau-Lifshitz-Gilbert-Slonczewski (LLGS) equation[34,35]

$$\frac{d\mathbf{m}}{dt} = -\frac{\gamma}{1+\alpha^2}\left(\mathbf{m}\times\mathbf{h}_{\text{eff}} + \alpha\mathbf{m}\times(\mathbf{m}\times\mathbf{h}_{\text{eff}})\right), \quad (1)$$

where $\mathbf{m} = \mathbf{M}/M_s$ is the normalized magnetization with $M_s$ corresponding to the saturation magnetization, $\gamma$ is the gyromagnetic ratio, and $\alpha$ the phenomenological Gilbert damping. Here $\mathbf{h}_{\text{eff}}$ is effective magnetic field,

$$\mathbf{h}_{\text{eff}} = \frac{2}{M_s}\left(A\nabla^2\mathbf{m} + D(\hat{\mathbf{z}}(\nabla\cdot\mathbf{m}) - \nabla m_z) + Km_z\hat{\mathbf{z}}\right), \quad (2)$$

where $A$, $D$, and $K$ are the strengths of the exchange, interfacial DMI and effective perpendicular anisotropy, respectively, and $\hat{\mathbf{z}}$ is the unit vector in the out-of-plane direction. The dipolar fields were included in the approximation of an infinite sample, $K = K_u - 0.5\mu_0 M_s^2$, where $K_u$ is the sample's easy-axis anisotropy and $\mu_0$ is the vacuum permeability. In the presence of VDMI and VCMA, we add to the effective field $\mathbf{h}_{\text{eff}}$ the following alternating fields,

$$\mathbf{h}_K = \Delta\mathrm{h}_K \cos(\omega t)\hat{\mathbf{z}} = \frac{2\Delta K}{M_s}\cos(\omega t)\hat{\mathbf{z}}, \quad (3a)$$

$$\mathbf{h}_D = \Delta\mathrm{h}_D \cos(\omega t)\hat{\mathbf{n}} = \frac{2\Delta D}{dM_s}\cos(\omega t)\hat{\mathbf{n}}, \quad (3b)$$

where $\Delta K$ and $\Delta D$ are the voltage-generated variations of the anisotropy and DMI, $d$ is the in-plane dimension of the lattice discretization, $\omega$ is the frequency, and $\hat{\mathbf{n}}$ is a unit vector in the x-y plane. The values of the parameters used in the micromagnetic simulations are given in Table 1.

| Parameter | Symbol | Value |
|---:|---|---|
| Exchange | $A$ | 20 pJ/m |
| Interfacial DMI | $D$ | (2.5 – 3.1) mJ/m² |
| Anisotropy | $K$ | 0.391 MJ/m³ |
| Magnetization Saturation | $M_s$ | $10^6$ A/m |
| Gilbert damping | $\alpha$ | 0.03 |
| Maximum Resistance | $R_{AP}$ | 1.5 kΩ |
| Minimum Resistance | $R_P$ | 1.0 kΩ |
| Isolated skyrmion sample | | $150\times150\times1$ with cell discretization of $0.5\times0.5\times1$ nm³ for $D = 2.5$ mJ/m² and cell discretization of $1\times1\times1$ nm³ for $D = 3.1$ mJ/m² |
| Single domain sample | | $100\times100\times1$ with cell discretization of $1\times1\times1$ nm³ |

**Table 1.** Parameters used in the micromagnetic simulations.

*Analytical model.* As ground state, we consider free-field stabilized isolated skyrmions, which can be described as circular domain walls, i.e. $\mathbf{m} = \sin\theta(r-R)(\cos(\psi+\eta)\hat{\mathbf{x}} + \sin(\psi+\eta)\hat{\mathbf{y}}) + \cos(r-R)\hat{\mathbf{z}}$ where $r$ and $\psi$ are the polar coordinates, and $R$ is the skyrmion radius.[36,37] This description is valid for $0.6 < D/D_c < 1$, where $D_c = 4\sqrt{AK}/\pi$ is the critical DMI, as already demonstrated.[8,37–39] The lowest order excitation of the skyrmion in the presence of homogenous fields and perturbations is the breathing mode, corresponding to a change in the skyrmion radius, which can be described in terms of the effective model,[37,40]

$$\frac{1}{\tau}\frac{d\eta}{dt} = \frac{\gamma}{M_s R}\left(A\left(-\frac{\Delta}{R^2}+\frac{1}{\Delta}\right)+K\Delta-\frac{\pi}{2}D\cos\eta\right)-\frac{\alpha}{\Delta}\frac{1}{\tau_\alpha}\frac{dR}{dt}, \quad (4a)$$

$$\frac{1}{\tau}\frac{dR}{dt} = -\frac{\pi\gamma}{2M_s}D\sin\eta+\alpha\Delta\frac{1}{\tau_\alpha}\frac{d\eta}{dt}. \quad (4b)$$

Here, $\Delta = \sqrt{A/K}$ is the domain wall width and the two parameters $\tau < 1$ and $\tau_\alpha < 1$ are associated to corrections to the skyrmion ansatz. The frequency of the breathing mode at linear order is,

$$\omega_s = \frac{4\pi\gamma K}{M_s}\left(1-\frac{D}{D_c}\right)\sqrt{\frac{8D}{D_c}}\tau, \quad (5)$$

which for the skyrmions considered, is lower than the ferromagnetic frequency,[37,39,41]

$$\omega_F = \frac{4\pi\gamma K}{M_s}. \quad (6)$$

The frequency from Eq. (5) is obtained by linearizing Eqs. 4(a)-(b), while the frequency from Eq. (6) is obtained by linearizing Eq. (1) in the absence of spatial gradients of the magnetization. However, the full Eqs. (1) and 4(a)-(b) are highly non-linear and a strong dependence of the resonance frequency with the amplitude is expected. Figure 2(a) shows a comparison between micromagnetic simulations and analytical calculations for the amplitude of the breathing mode, in terms of the skyrmion radius ΔR, as a function of the frequency of the applied field $\Delta h_K$ originated from the VCMA. We consider a skyrmion stabilized with $D$ = 3.1 mJ/m² and average radius around 15 nm. We observe that, for small $\Delta h_K$, the characteristic gaussian curve of a linear excitation is obtained. As $\Delta h_K$ increases, the profile modifies, revealing a nonlinear behavior, with a frequency shift and a profile asymmetry around the highest amplitude. A similar behavior is observed for high power excitation of uniform magnetic states.[42] For the skyrmion configuration, this non-linear behavior is due to the asymmetric nonlinear potential (see Supplemental Material). We also notice that for higher amplitudes of the breathing mode, when the amplitude of the excitation ΔR is close to the skyrmion radius, the model in Eqs. 4(a)-(b) undervalues the amplitude due to the nonlinear behavior, as discussed in the literature.[37] The qualitative agreement shows a resemblance to the behavior of a Duffing oscillator.[43,44] We see for example, a strong dependence of the resonance frequency with the amplitude of the applied field. A remark, however, is the lack of a peak at twice the resonant frequency, which evidences the inefficiency of parametric excitation.[14]

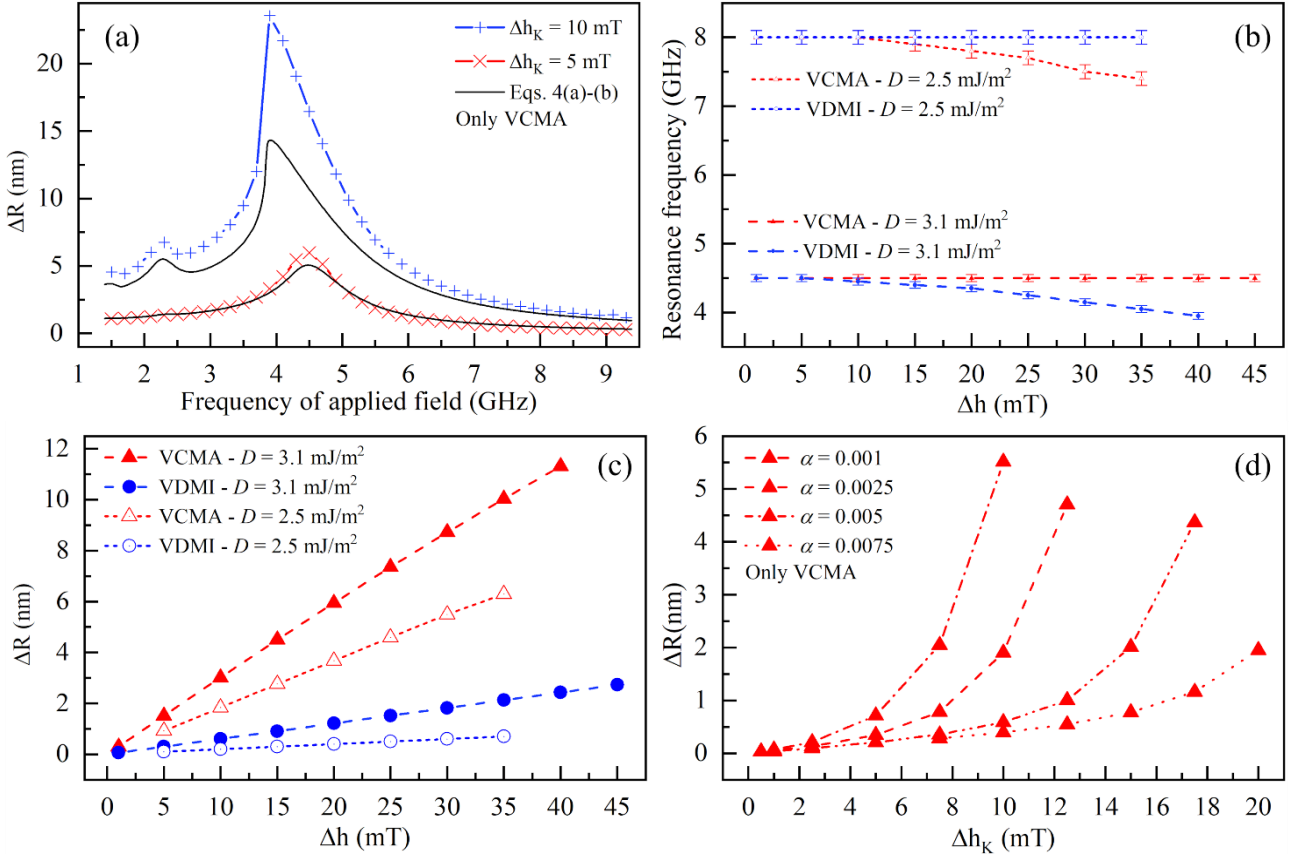

**FIG. 2.** (a) A comparison of the micromagnetic and analytic amplitude of the breathing mode in term of skyrmion radius (ΔR) as a function of the microwave frequency of the applied field Δh$_K$ originated from the VCMA. Micromagnetic simulation results of: (b) Resonance frequency as a function of the amplitude of the applied field (Δh) originated from VCMA or VDMI. (c) Amplitude of the breathing mode in term of skyrmion radius (ΔR) as a function of the applied field (Δh) originated from VCMA or VDMI. (d) Amplitude of the breathing mode in term of skyrmion radius for different damping values, considering $D = 3.1 mJ/m^2$ and frequency of the applied field Δh$_K$ as $2\omega_s$, i.e., double the resonance frequency.

To investigate further the nonlinear behavior, we performed micromagnetic simulations to evaluate the resonance frequency as a function of the amplitude of the applied field originated from the VCMA or the VDMI, shown in Fig. 2(b). For larger amplitudes, the frequency decreases almost linearly with the applied field. Here, we considered two skyrmions: one stabilized with $D = 3.1$ mJ/m$^2$ and average radius around 15 nm, and one stabilized with $D = 2.5$ mJ/m$^2$ and average radius around 7 nm. Similar qualitative results are obtained for $D > 2.1$ mJ/m$^2$, below this value no skyrmion was stabilized.

Figure 2(b) shows that the VCMA is more efficient in driving the nonlinear resonance than VDMI. This is associated to a larger excitation amplitude induced by the variation of the anisotropy parameter, as we can understand from the analytical model in Eqs. 4(a)-(b). We see that reparametrizing the spatial distances by the domain wall width, and the energy by the domain wall energy, reveals that the skyrmion stability and dynamics depends only on a single dimensionless parameter $g = \pi D/4\sqrt{AK}$. Hence, the relative excitation efficiency of VDMI and VCMA depends on the voltage-generated variations $\Delta D$ and $\Delta K$ as well as on the material parameters $D$ and $K$. Specifically, changes in the dynamical parameter by $\Delta h_K$ and $\Delta h_D$ give a ratio of $\delta g_D/\delta g_K = -2(Kd/D)(\Delta h_D/\Delta h_K)$. For the values used in the simulations, $2(Kd/D)$ is around 0.16 and 0.25 for $D = 2.5$ mJ/m$^2$ and $D = 3.1$ mJ/m$^2$, respectively, implying that the variation of $g$ (see Supplemental Material) due to $\Delta D$ is much smaller than by $\Delta K$. We also notice that the material parameters are strongly related to the skyrmion size, such that smaller $D$ values are associated to smaller skyrmions. Thus, smaller skyrmions are more strongly affected by VCMA and VDMI. These results are corroborated by micromagnetic simulations. Figure 2(c) shows the results of micromagnetic simulations analysing the amplitude of the breathing mode at different applied fields for the two different values of DMI. We notice that, indeed, the efficiency $\Delta R/\Delta h$ of the VCMA and VDMI is strongly affected by the DMI. Moreover, the same voltage excites dynamical changes of the skyrmion radius with a phase difference of $\pi/2$ for the different mechanisms. The phase difference can also be understood from the fact that, while the DMI couples with the gradient of **m**, the anisotropy couples directly with **m**, see Eq. 2.

As a complementary analysis, we also investigated the parametric excitation of the skyrmion with twice the resonant frequency. Figure 2(d) shows the amplitude of the excited mode with a change in the amplitude of the applied field at twice the resonant frequency. Although we notice the expected quadratic growth, it is strongly damped even at extremely low values of $\alpha$. These results agree with the effective theory Eqs. 4(a)-(b). Parametric resonance is often expected in non-linear systems, but it is not guaranteed.[45] It depends heavily on the non-linear potential. It has been previously demonstrated for magnonic systems in collinear backgrounds[46–50] and predicted in antiferromagnets[51]. For a skyrmion, however, there is a strong asymmetry of the potential around the equilibrium radius, which leads to a shift in the average radius for higher excitations[37] (see Supplemental Material). Numerical calculations show that this hinders the parametric excitation. These results are crucial for the design of skyrmion-based technologies in a proper working regime and guides future experimental efforts concerning the excitation of skyrmions.

*Sensitivity of spintronic diode based on VCMA and VDMI.* To reduce the power dissipation, i.e., Joule heating, of SDs, the excitation should be driven by voltage controlled magnetic parameters. To verify its efficiency, Fig. 3(a) compares the excitation of the skyrmion breathing mode at the resonant frequency, see Eq. (5), and the ferromagnetic resonance of a uniform magnetic state from Eq. (6). We plotted the variation of the out-of-plane component of the magnetization, which gives rise to the resistance variation of the device, considering a sample with an isolated skyrmion at $D = 2.5$ mJ/m$^2$ and $D = 3.1$ mJ/m$^2$, and a uniform domain at $D = 3.1$ mJ/m$^2$. We emphasize that the sensitivity of the skyrmion-based device should be higher for smaller $D$. The resonant frequency is also higher for smaller $D$, which allows for the tunability of the device and a broader range of operational frequencies. To consider an unconstrained skyrmion, we take into account a square sample with size around five times the skyrmion radius. For the single domain device, the excitation by VCMA is invariant of the area of the sample, while the VDMI only couples to the magnetization gradient generated at the edges due to boundary conditions[38]. Therefore, we consider a sample for the uniform magnetization as a square area of $100$ nm $\times 100$ nm.

As expected, skyrmions exhibit a larger variation of the out-of-plane magnetization component due to VCMA and VDMI, while the single domain is almost invariant under variation of the VDMI. In particular, the smaller skyrmion, with lower $D$, presents the larger size variation to the applied VCMA and VDMI. This advocates for the scalability of the skyrmion-based SD. At the largest calculated variation of the magnetization, the skyrmion shows an amplitude 10 times bigger than that of a single domain, which promises a large sensitivity for skyrmion-based SDs. To verify this, we calculate the sensitivity as a function of the applied voltage in Fig. 3(b).

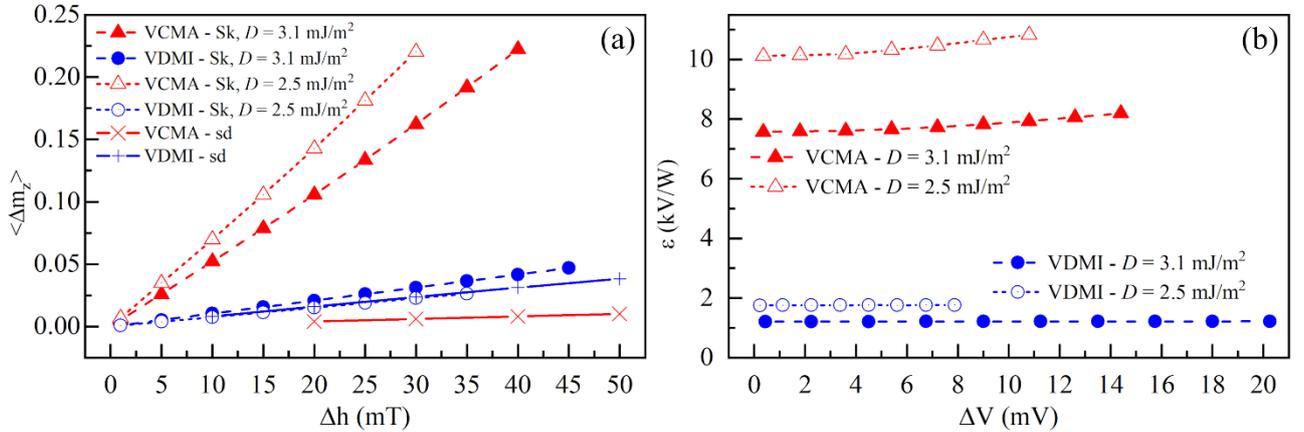

**FIG. 3.** (a) Amplitude of the dynamical average out-of-plane component f the magnetization <Δm$_z$> as a function of the applied field Δh originated from VCMA or VDMI. We compare the excitation of a skyrmion (Sk) and single domain (sd). (b) Sensitivity as a function of the applied voltage for a skyrmion-based SD.

For the calculation of the predicted sensitivity, we consider the resistance $R_{MTJ}$ as a function of the total $m_z$ component,

$$R_{MTJ} = R_P + 0.5(R_{AP} - R_P)(m_z - 1). \tag{6}$$

Based on the resistance variation and the VDMI efficiency of 1 pJ/Vm,[24,25] we obtain the input power $P_{in}$ and sensitivity $\varepsilon$ of the SD given by

$$P_{in} = \frac{V_{eff}^2}{R_{avg} + R_I}, \quad (7a)$$

$$\varepsilon = \frac{R_{avg} I_{avg}}{P}, \quad (7b)$$

where, $V_{eff}$ is the average effective applied voltage, $R_{avg}$ and $R_I$ are the average and characteristic impedance resistances of the device respectively, and $I_{avg}$ is the effective current through the device due to the applied $V_{eff}$. Figure 3(b) shows the sensitivity of the skyrmion-based SD as a function of the applied voltage considering excitation by VCMA and VDMI for an isolated skyrmion at $D = 2.5$ mJ/m$^2$ and $D = 3.1$ mJ/m$^2$. The sensitivity is about rather invariant of the applied voltage and reaches the value of 10 kV/W for the smallest considered skyrmion, and is five times higher the previously predicted for active SDs assisted by the VCMA[8]. As a comparison, the highest sensitivity obtained for the single domain with the data of Fig. 3(a) is 420 V/W, around 23 times smaller than the one obtained in the presence of the skyrmion. We also remark that, the currents through the SDs are around 0.5 µA, while previous studies have required currents of around 50 µA[8], which implies a much lower Joule dissipation for the SDs thanks to the larger MgO barrier necessary to achieve large voltage-control of physical parameters.

*Summary and conclusions.* Here, we explored the implementation of voltage-assisted skyrmion-based SDs. We performed a systematic study of the skyrmion excitation by both VCMA and VDMI. The main result is that VCMA is more efficient than VDMI, which leads to a sensitivity for the skyrmion-based diode larger than 10kV/W. We showed that it is much larger than the sensitivity obtained by a single domain in the same configuration and than previous results reported in the literature[8]. We also emphasize that smaller values of DMI, corresponding to smaller skyrmions, showed a higher sensitivity opening a path for the scalability of skyrmion-based SDs. For larger excitation, we reported a non-linear behavior, characterized by a shift of the frequency as well as a shift of the average skyrmion radius. Moreover, we noticed that skyrmions, due to the non-linearity and asymmetry of their energy landscape, cannot be effectively excited by twice the frequency within state-of-art experimental conditions, unlike usually expected. This work combined to current efforts to improve anisotropy and DMI control paves the way for ultra-sensitive and low power skyrmionics devices.

ACKNOWLEDGEMENT

This work was supported under the project number 101070287 — SWAN-on-chip — HORIZON-CL4-2021-DIGITAL-EMERGING-01, the project PRIN 2020LWPKH7 "The Italian factory of micromagnetic modelling and spintronics", and D.M. 10/08/2021 n. 1062 (PON Ricerca e Innovazione) funded by the Italian Ministry of University and Research, and by PETASPIN association (www.petaspin.com). The work of GF has been partially funded by European Union (NextGeneration EU), through the MUR-PNRR project SAMOTHRACE (ECS00000022). The work of DR, RT, MC has been partially funded by European Union (NextGeneration EU), through the MUR-PNRR project NEST (PE0000021).